\documentstyle[12pt,worldsci,epsf]{article}

\newfont{\bg}{cmr10 scaled\magstep4}

\newcommand{\bigzerou}{\smash{\lower1.ex\hbox{\bg 0}}}

\begin{document}
\title{
Monopole Condensation in Lattice SU(2) QCD
\footnote{
This is based on works done in collaboration with Y.Matsubara, H.Shiba,
S.Kitahara, S.Ejiri, S.Okude and K.Yotsuji.
}
}
\author{
 Tsuneo Suzuki
\thanks{ E-mail address: suzuki@hep.s.kanazawa-u.ac.jp}
\\
{\em Department of Physics, Kanazawa University, Kanazawa 920-11, Japan
}}

\maketitle

\begin{abstract}
This is the short review of 
 Monte-Carlo studies of quark 
confinement in lattice QCD. After 
abelian projections both in the maximally abelian and Polyakov gauges, 
it is seen that 
the monopole part alone is responsible for confinement.
A block spin transformation on the dual lattice 
 suggests 
that lattice $SU(2)$ QCD is always ( for all $\beta$) in the monopole 
condensed phase and so in the confinement phase in the infinite volume 
limit. 
\end{abstract}

\vspace{1cm}

\section{Introduction}

It is crucial to understand the mechanism of quark confinement 
in order to explain hadron physics out of QCD.
Our standpoint is based on the 'tHooft idea of abelian projection 
of QCD.\cite{thooft}
The abelian projection is to fix the gauge in such a way that the maximal 
torus group remains unbroken.
After the abelian projection, monopoles appear as a topological quantity 
in the residual abelian channel.
QCD is reduced to an abelian theory with electric charges and monopoles.
If the monopoles make Bose condensation, charged quarks and gluons 
are confined due to the dual Meissner effect. 

Based on this standpoint, the present author and his collaborators
 have studied color 
confinement mechanism and hadron physics performing
Monte Carlo simulations of abelian projection 
in lattice QCD.
\cite{yotsu,suzu93,shiba3,shiba4,shiba5,shiba6,ejiri2,ejiri3,suzu94a,suzu94b,matsu,kita94,suzu94c,suzu95a}
The aim of the study is to ascetain correctness of 
the picture, that is, to check if monopole condensation really 
occurs in QCD.
Here I  review the results of 
these studies compactly.\footnote{
We have also studied hadron physics based on 
an infrared effective Lagrangian constructed 
directly from QCD on the assumption of the above picture
\cite{suzu90,matsu91,monden,kamizawa,kodama}, but this time
I skip all of them.}

\section{Abelian dominance and monopole dominance}
\subsection{Maximally abelian and Polyakov gauges}
There are infinite ways of abelian projection extracting such an 
abelian theory out of  QCD. 
We have found two gauges which show interesting behaviors 
called abelian dominance.
\cite{yotsu,suzu93,kron}

One is the maximally abelian (MA) gauge.
Define a matrix in $SU(2)$ QCD
\begin{eqnarray*}
X(s) & = & \sum_{\mu}[
           U(s,\mu)\sigma_3 U^{\dagger}(s,\mu) 
      + U^{\dagger}(s-\hat{\mu},\mu)\sigma_3 U(s-\hat{\mu},\mu)] \\
     & = & X_1 (s)\sigma_1 + X_2 (s)\sigma_2 + X_3 (s)\sigma_3.
\end{eqnarray*}
Then a gauge satisfying $X_1 (s)=X_2 (s)=0$ is the MA gauge
\cite{kron}
which tends to a $U(1)$-covariant gauge $(\partial_{\mu} \pm igA^3_{\mu})
A^{\pm \mu} =0$ in the continuum limit.
The other is the Polyakov gauge 
which is defined by diagonalizing Polyakov loop operators. 
We have found that 
important features of confinement, i.e., the string tension and the 
charateristic behaviors of the Polyakov loops are well reproduced in 
terms of the abelian operators $O(u(s,\mu))$ both 
in MA gauge\cite{yotsu,suzu93} and in the Polyakov gauge.
\cite{suzu94a} 
See Figures\  $1 \sim 4$.

\begin{figure}[t]
\begin{flushleft}
\epsfxsize=6cm
\leavevmode
\epsfbox{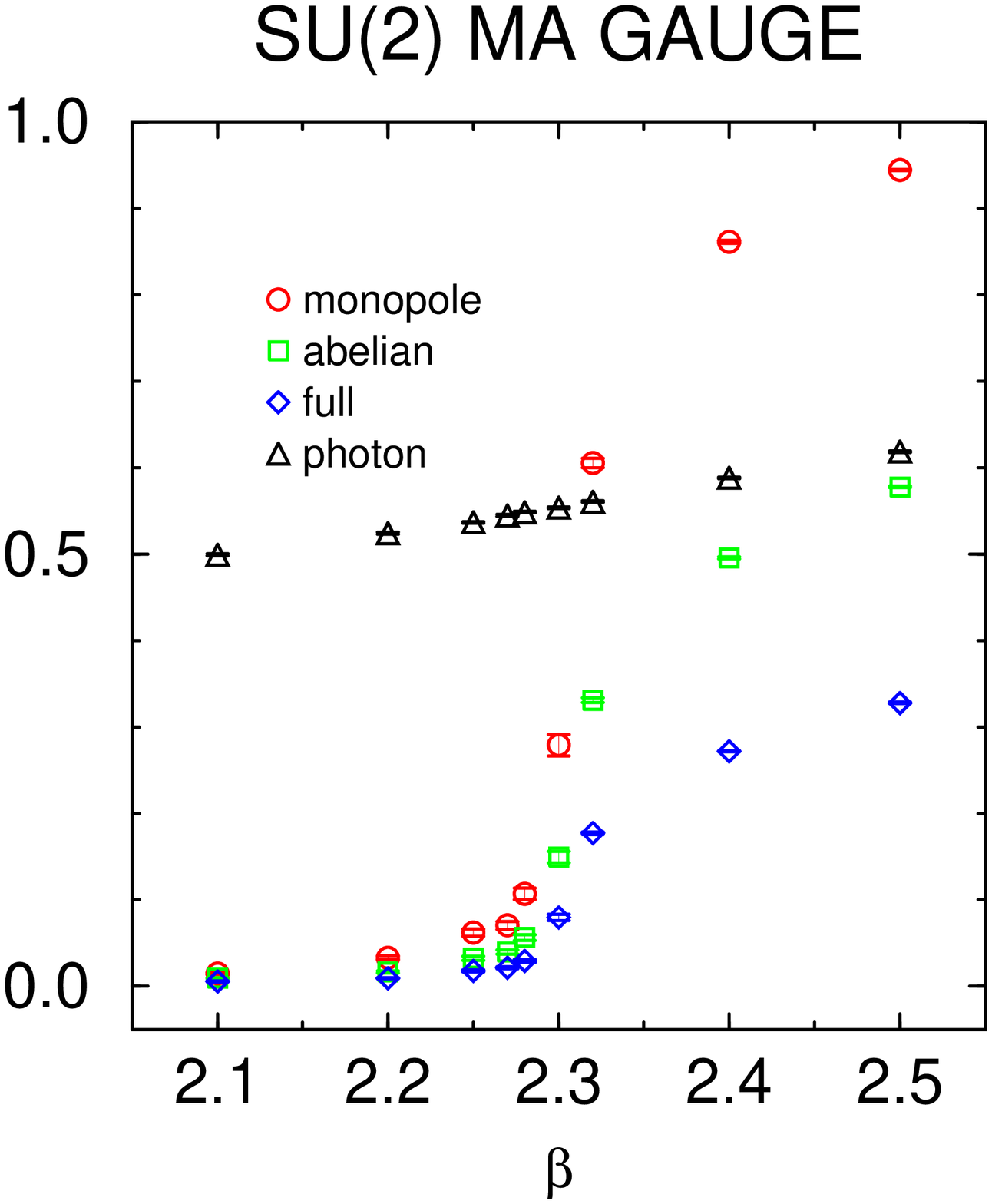}
\end{flushleft}
\begin{flushleft}
\vspace{-1cm}
Fig.1 \ Monopole Dirac string\\
and photon contributions to \\
Polyakov loops in MA gauge.
\end{flushleft}
\label{f1}

\vspace{-10.3cm}

\begin{flushright}
\epsfxsize=6cm
\leavevmode
\epsfbox{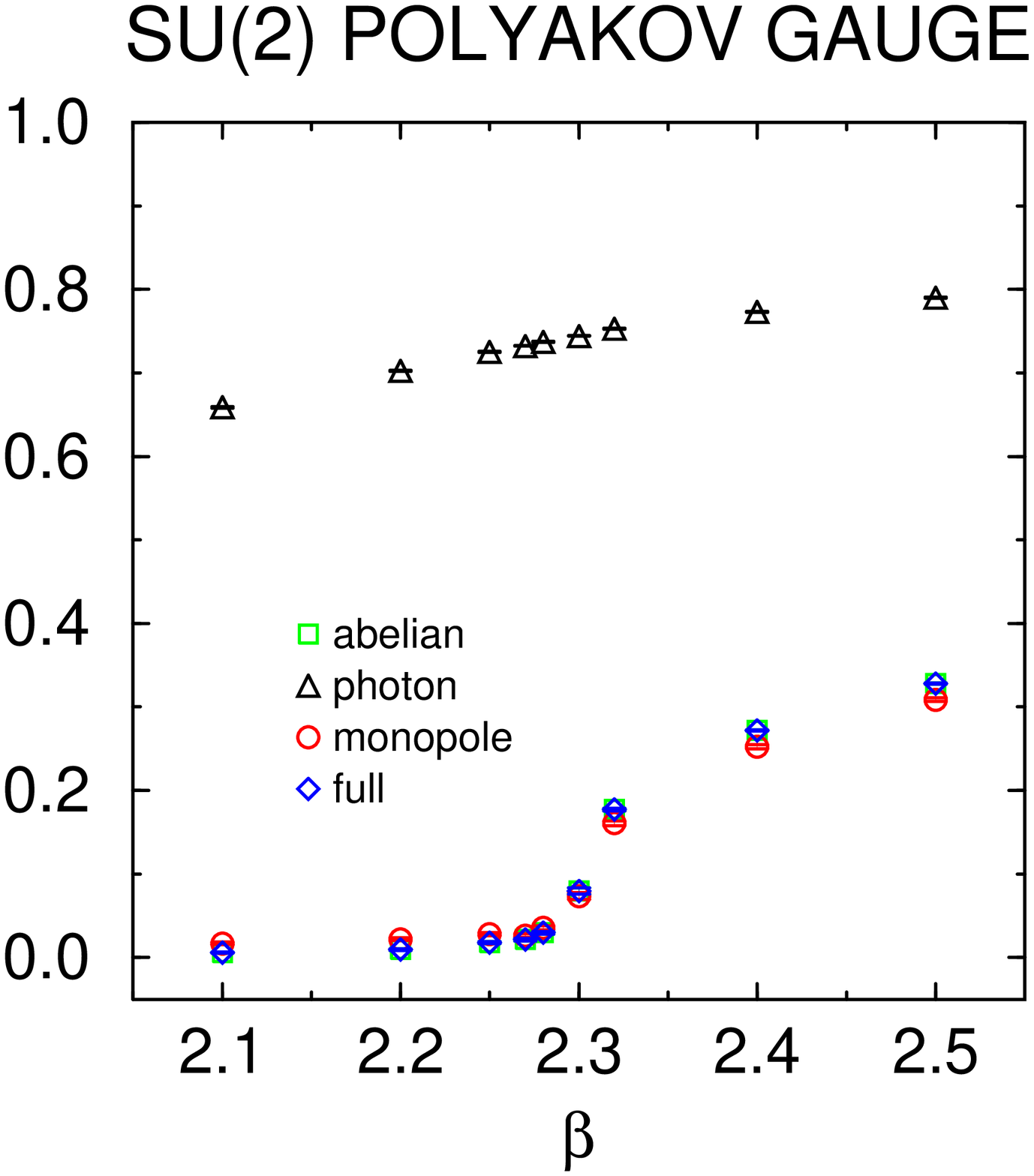}
\vspace{-1cm}
\end{flushright}
\begin{flushright}
\parbox{6cm}{
\begin{flushleft}
Fig.2 \ Monopole Dirac string
and photon contributions to 
Polyakov loops in the 
Polyakov gauge.
\end{flushleft}
}
\end{flushright}
\label{f2}
\end{figure}


\subsection{Monopole dominance in the Polyakov loops}
An abelian Polyakov loop $P$ which is written in terms of abelian link fields 
alone 
is given by a product of contributions from Dirac strings 
of monopoles and from photons.\cite{suzu94a,matsu} 
We have observed the photon and the Dirac-string 
contributions separately. See Figs.1 and 2.
The characteristic features of the Polyakov loops as an order 
parameter of the deconfinement transition are 
due to the Dirac string contributions alone.
The photon part has a finite contribution for both phases and 
it changes only slightly.
The fact that monopoles are responsible for the essential feature of the 
Polyakov loop is found also in $U(1)$ and in $SU(3)$.
It is remarkable that
the behavior of the abelian Polyakov loops as an order parameter and the 
monopole responsibility are seen 
{\it in any gauge.\cite{suzu94a}}
This is the first phenomena suggesting gauge independence of 
the 'tHooft conjecture.


\begin{figure}[tb]
\epsfxsize=6cm
\begin{flushleft}
\leavevmode
\epsfbox{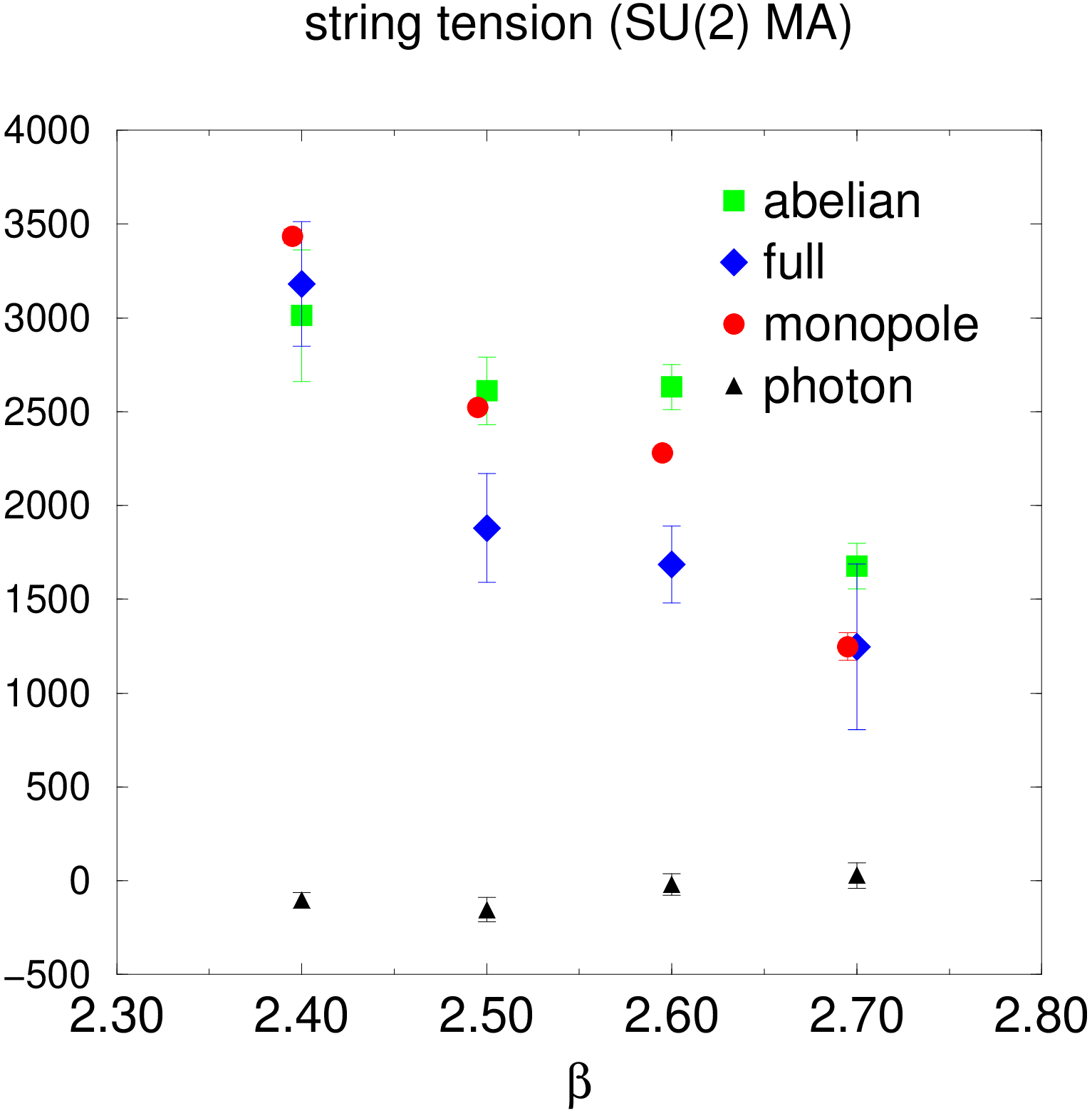}
\end{flushleft}
\begin{flushleft}
\vspace{-3cm}
Fig.3\  Monopole and photon \\
contributions to the string \\
tension in MA gauge in $SU(2)$\\
QCD in comparison with the full\\ 
and the abelian ones.
\end{flushleft}
\label{f3}

\vspace{-9.6cm}

\epsfxsize=6cm
\begin{flushright}
\leavevmode
\epsfbox{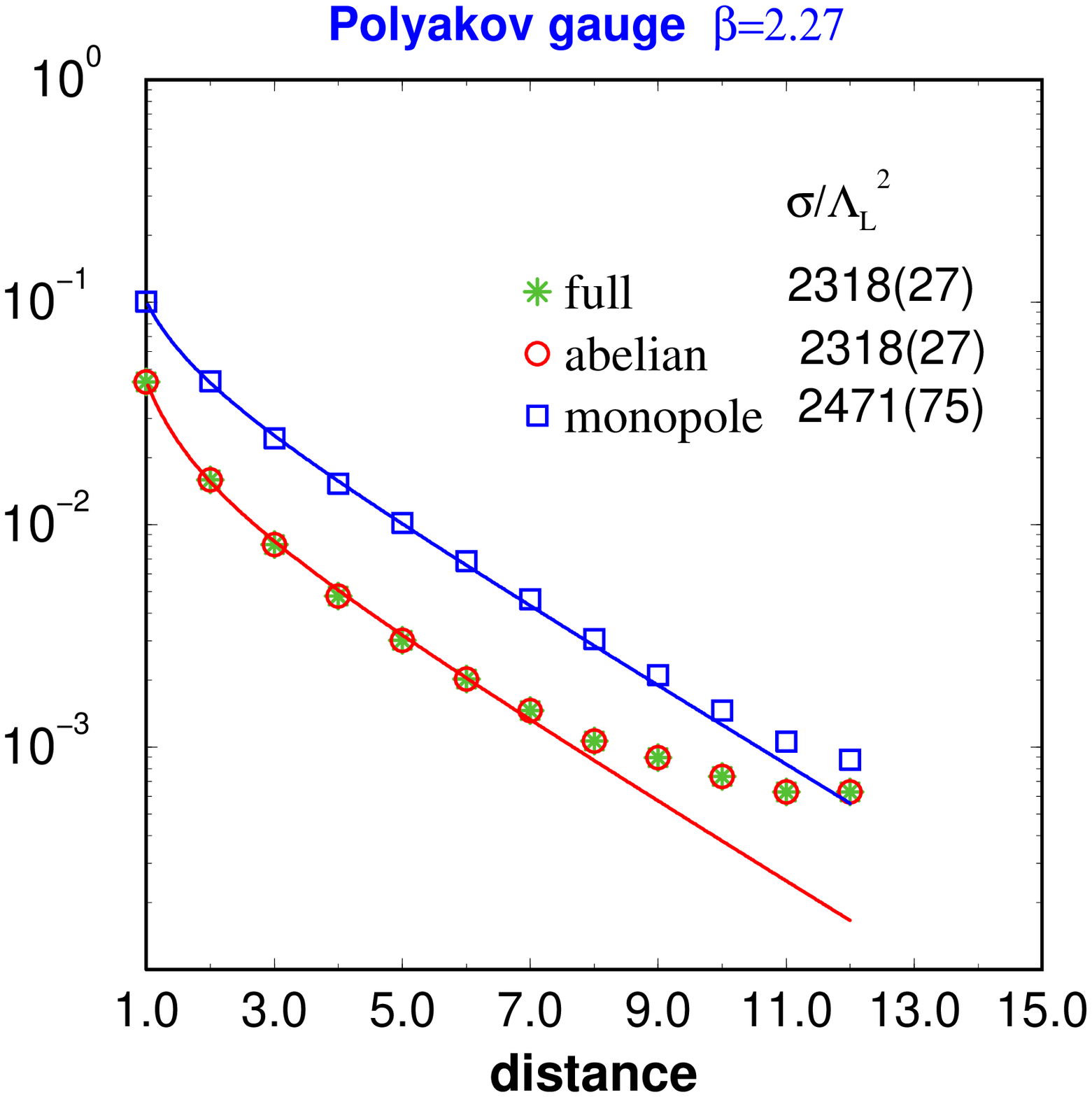}
\end{flushright}
\vspace{-2cm}
\begin{flushright}
\parbox{6cm}{
\begin{flushleft}
Fig.4 \ Polyakov loop correlations
in the Polyakov gauge. 
\end{flushleft}
}
\end{flushright}
\label{f4}
\vspace{1cm}
\end{figure}


\subsection{The string tension and monopoles}
Shiba and Suzuki\cite{shiba3,shiba4} have shown furthermore 
that monopoles alone can reproduce the full 
value of the string tension in $SU(2)$ QCD.
An abelian Wilson loop operator 
is rewritten  by a product of monopole and photon contributions. 
The string tensions evaluated from monopole and photon contributions 
are plotted in Fig.3. The string tension is reproduced only from 
the monopole contribution.

It is possible to know the static potential from the correlation of the 
Polyakov loops. The string tension is then derived from the static potential.
We have measured the correlations using the Polyakov 
loop operators written in 
terms of full and abelian link fields and link fields given by the 
monopole Dirac string.
The preliminary data are plotted in Fig.4 in the case of the Polyakov
gauge. A linear behavior is seen with the almost similar string tension
in all three cases. This is another data supporting gauge invariance of 
confinement mechanism due to monopole condensation.

\section{Monopole  action and condensation}
The above abelian dominance suggests that a set of $U(1)$ invariant operators 
$\{O(u(s,\mu))\}$ are enough to describe confinement. 
Then there must exists an effective $U(1)$ action 
$S_{eff}(u)$ describing confinement.
We tried to derive $S_{eff}(u)$ using Schwinger-Dyson equations, but failed 
to get it in a compact and local form.\cite{suzu93}
$S_{eff}(u)$ contains larger and larger loops as $\beta$.

Shiba and Suzuki author tried to 
perform a dual transformation of $S_{eff}(u)$ in $SU(2)$ QCD and to obtain 
the effective $U(1)$ action 
\cite{shiba3,shiba6,suzu94b}
in terms of monopole currents.\cite{degrand}
To study the long range behavior important in QCD, They have considered also
extended monopoles.\cite{ivanenko} 
The extended monopole currents are defined by the number of the Dirac strings
 surrounding an extended cube:
\begin{eqnarray}
k_{\mu}^{(n)}(s) 
    & = & \sum_{i,j,l=0}^{n-1}k_{\mu}(ns+(n-1)\hat{\mu}+i\hat{\nu}
     +j\hat{\rho}+l\hat{\sigma}),
\end{eqnarray}
where $k_{\mu}(s)$ is the ordinary monopole current.\cite{degrand}
Considering extended monopoles corresponds to performing a block spin 
transformation on a dual lattice\cite{shiba6,suzu94b} and so It is 
suitable for exploring the long range property of QCD.

The partition function of interacting monopole currents is expressed as
\begin{eqnarray}
  Z=(\prod_{s,\mu} \sum_{k_{\mu} (s)=-\infty}^\infty ) \,
    (\prod_{s} \delta_{\partial '_{\mu}k_{\mu}(s) ,0 } ) \,
    \exp (-S[k]) .
\label{eqn:pfunm}
\end{eqnarray}
It is natural to assume $S[k] = \sum_i f_i S_i [k]$. Here
$f_i$ is a coupling constant of an interaction $S_i [k]$. 
For example, $f_1$ is the coupling of the self energy term  
$\sum_{n,\mu}(k_\mu(s))^2$, $f_2$ is the coupling 
of a nearest-neighbor interaction term
 $\sum_{n,\mu} k_\mu(s) k_\mu(s+\hat\mu)$ 
and $f_3$ is the coupling of another nearest-neighbor term 
$\sum_{n,\mu\neq\nu} k_\mu(s) k_\mu(s+\hat\nu)$.\cite{shiba6}
Shiba and Suzuki\cite{shiba3,shiba5,shiba6} extended a method
developed by Swendsen\cite{swendsn} to the system of monopole currents 
obeying the current conservation rule.

The monopole actions are obtained locally enough 
for all extended monopoles considered even in the scaling region.
They are 
lattice volume independent.
The coupling constant $f_1$ of the self-energy term is dominant 
and the coupling constants decrease 
rapidly as the distance between the two monopole currents increases.

To study monopole dynamics, we have also studied the length of monopole loops
 and monopole charge distribution.
We have found that 
1)each vacuum has only one long connected loop and a small number of short
loops. 
The difference of the two types of loops is very clear.
2)The length of the long loop becomes shorter as $\beta$ becomes larger. 
In the deep deconfinement region, there disappears a long loop.
3)Monopole charges are almost $\pm 1$.

\begin{figure}[t]
\epsfxsize=6cm
\begin{flushleft}
\leavevmode
\epsfbox{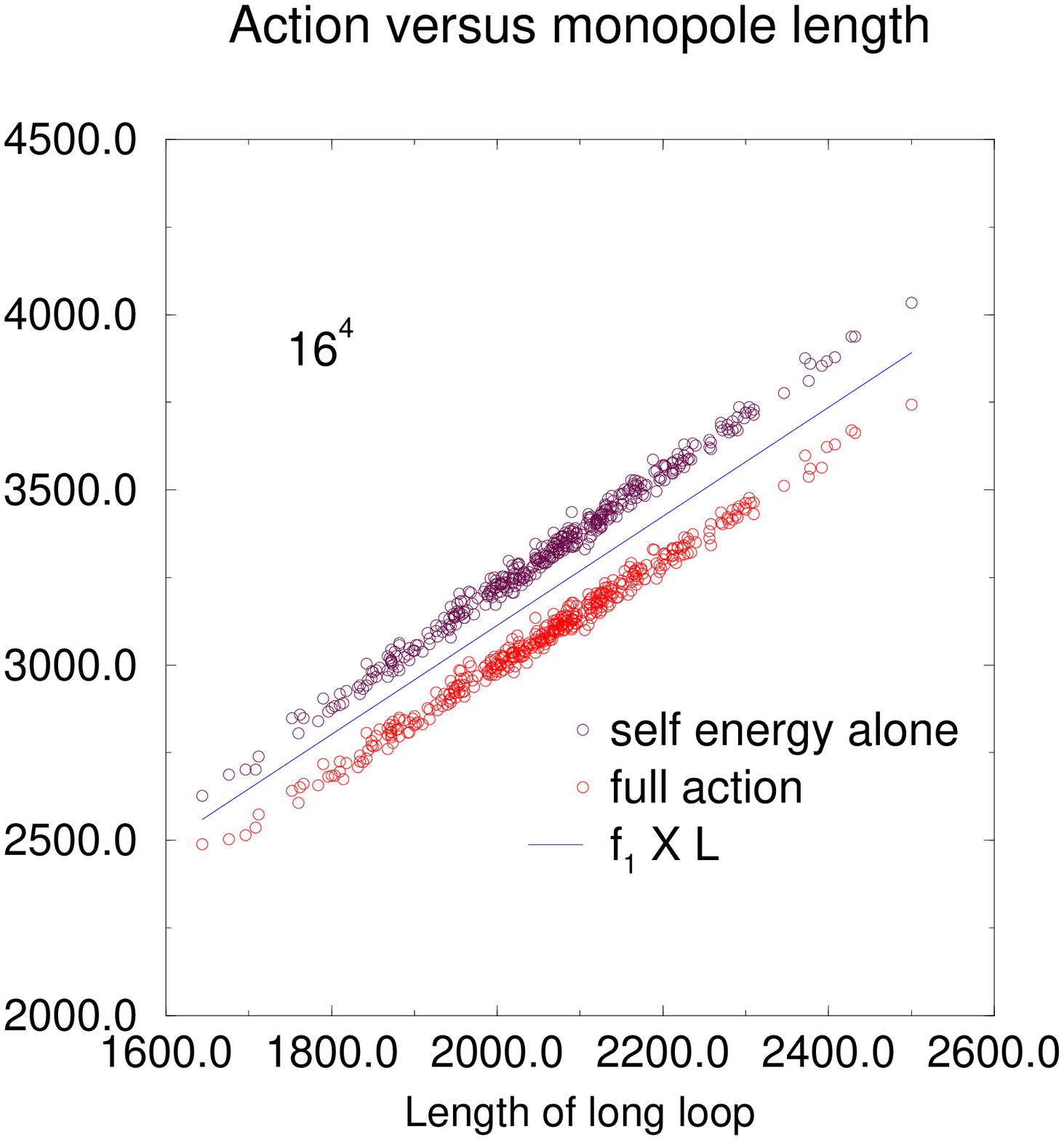}
\end{flushleft}
\vspace{-1cm}
\begin{flushleft}
Fig.5 \ Monopole action versus\\
 length of loops
\end{flushleft}
\label{f7}

\vspace{-10cm}

\begin{flushright}
\epsfxsize=6cm
\leavevmode
\epsfbox{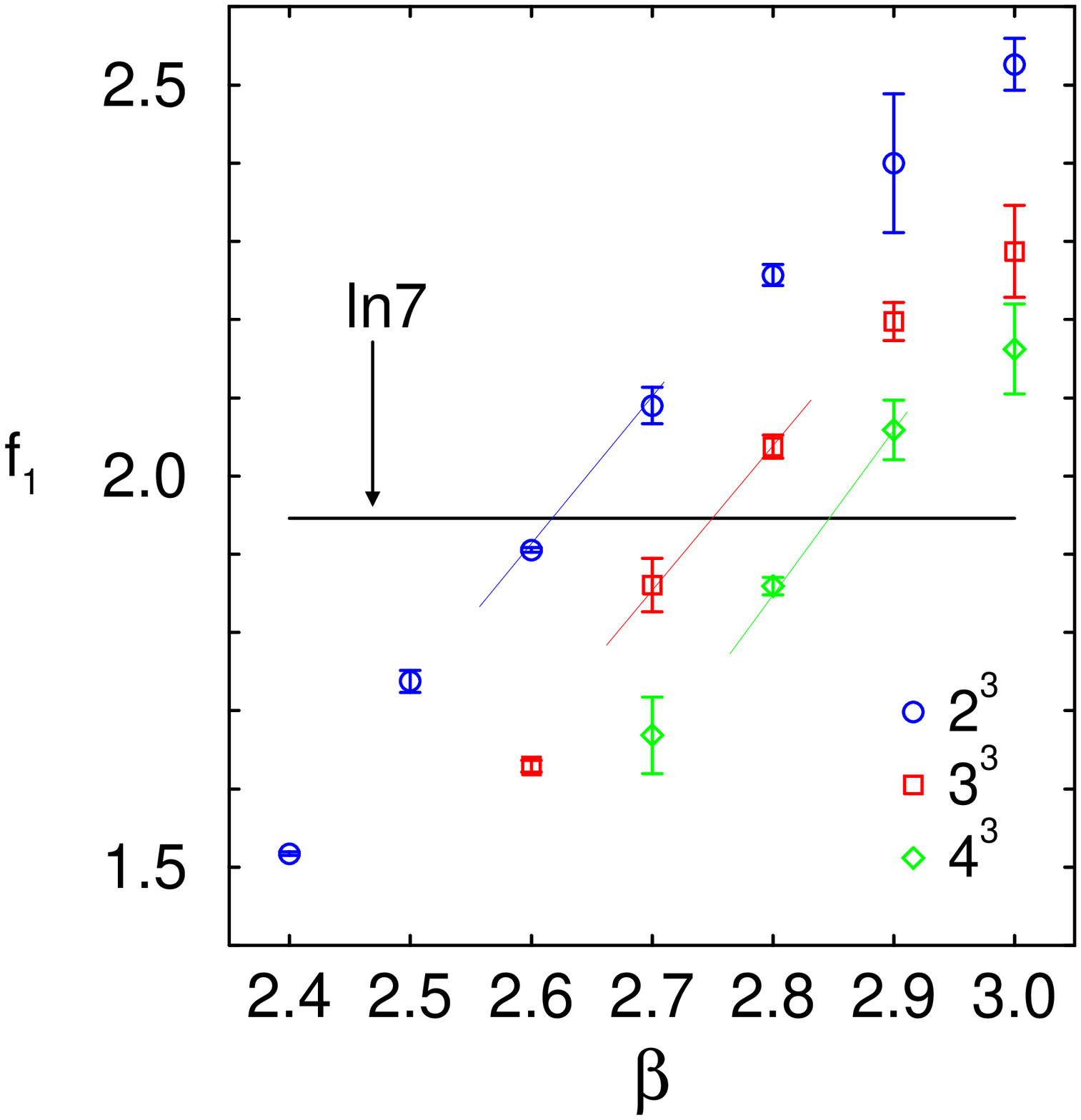}
\end{flushright}
\vspace{-1.5cm}
\begin{flushright}
\parbox{6cm}{
\begin{flushleft}
Fig.6\  Coupling constants $f_1$ versus
 $\beta$ for $2^3,3^3$, and $4^3$ 
monopoles  on $24^4$ lattice.
\end{flushleft}
}
\end{flushright}
\label{f8}
\end{figure}

Since the long loop is important, we plot the value of 
monopole action versus length of the long loops as seen in Fig.5.
The value of the action is proportional to the length $L$ of the long loop
is well approximated by $f_1 \times L$. 

As done in compact QED,\cite{bank} 
the entropy of a monopole loop can be estimated as $\ln 7$ 
per unit loop length.
Since
the action is approximated by 
the self energy part $f_1 L$,
the free energy per unit monopole loop length is approximated by 
 $( f_1-\ln 7 ) $ .
If $f_1 < \ln 7$, the entropy dominates over the energy, 
which means condensation of monopoles.
In Fig.6, $f_1$ versus $\beta$ for various extended 
monopoles on $24^4$ 
lattice is shown in comparison with the entropy value $\ln 7$.
Each extended monopole has its own $\beta$ region where the condition
$f_1 < \ln 7$ is satisfied.
When the extendedness is bigger, larger $\beta$ is included in such a 
region. Larger extended monopoles are more important in determining 
the phase transition point.

\begin{figure}[t]
\epsfxsize=6cm
\begin{flushleft}
\leavevmode
\epsfbox{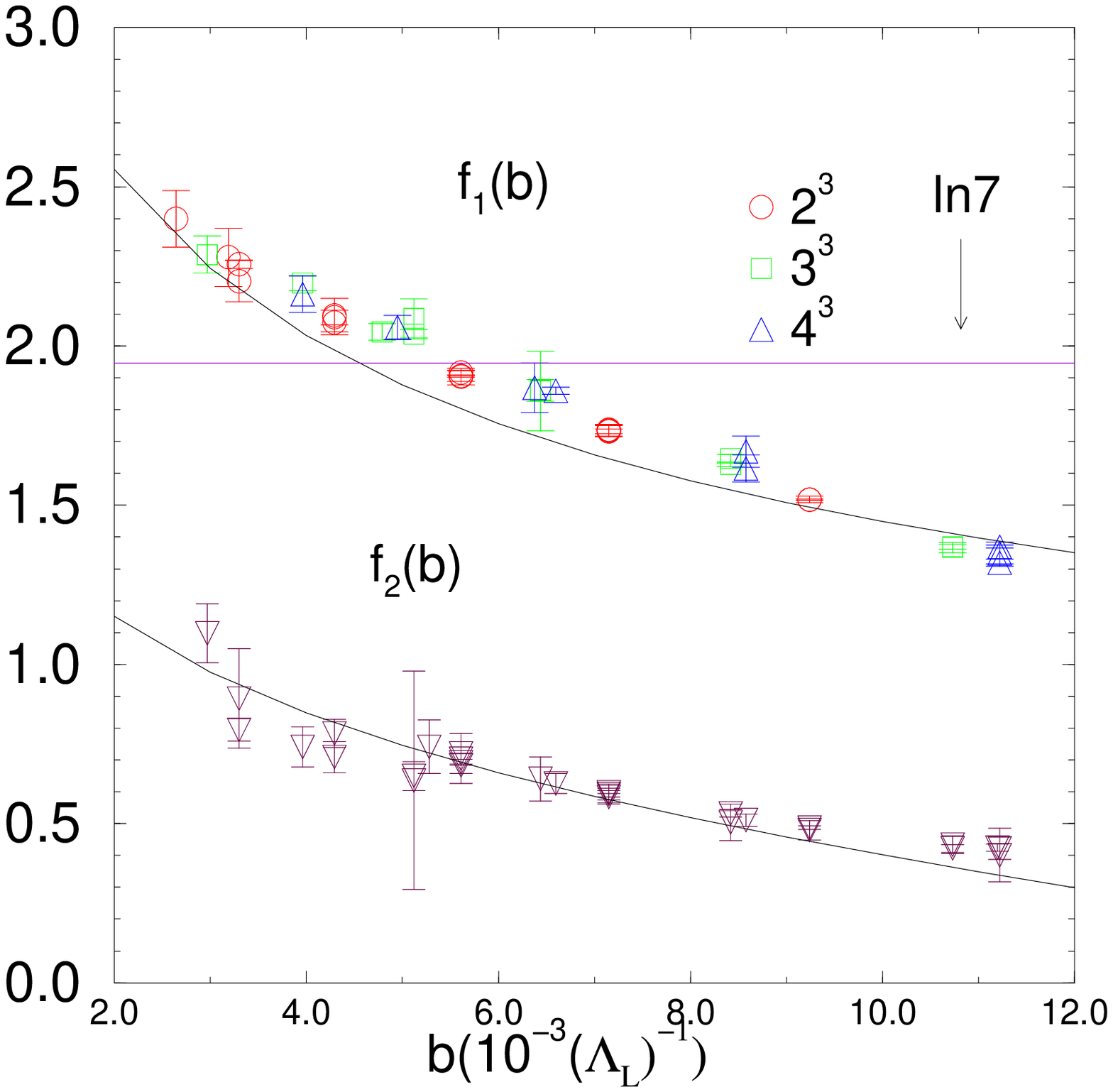}
\end{flushleft}
\vspace{-2cm}
\begin{flushleft}
Fig.7\ Coupling constants $f_1$ \\
and  $f_2$ versus $b$.
\end{flushleft}
\label{f9}

\vspace{-9.5cm}

\epsfxsize=6cm
\begin{flushright}
\leavevmode
\epsfbox{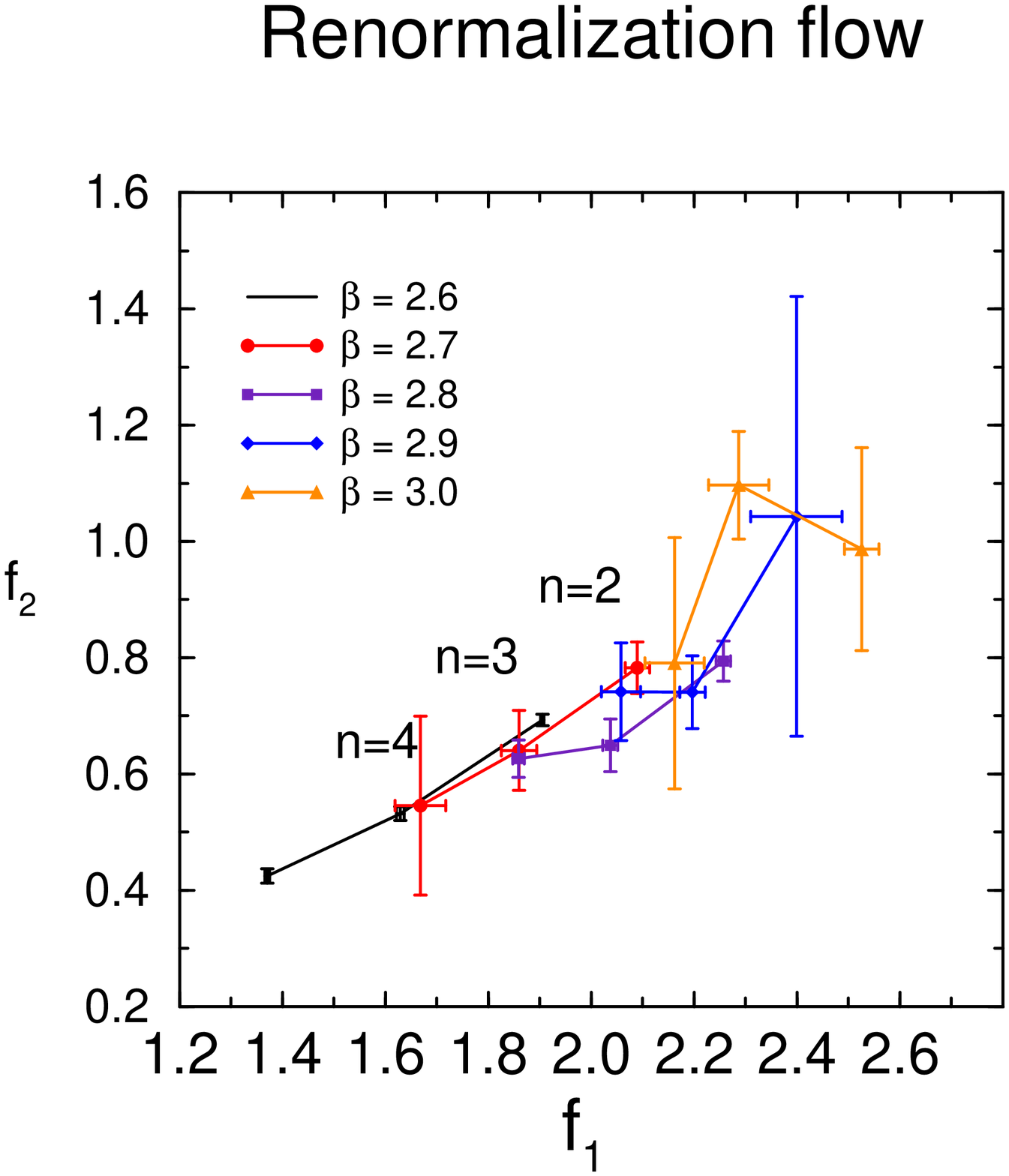}
\end{flushright}
\vspace{-2cm}
\begin{flushright}
\parbox{6cm}{
\begin{flushleft}
Fig.8\ Renormalization flow in the $f_1-f_2$ plane
\end{flushleft}
}
\end{flushright}
\label{f10}
\end{figure}

The behaviors of $f_i$ are 
different for different extended 
monopoles. However, 
if we plot them versus 
$b=n\times a(\beta)$,
we get a unique curve as in Fig.7. The 
coupling constants seem to depend only on $b$, not on the extendedness 
nor $\beta$.  
There is a critical $b_c$ corresponding to critical $\beta^n_c$,
 i.e., $b_c =na(\beta^n_c)$.

Now we can derive important conclusions.
Suppose the effective monopole action remains the same for any 
extended
 monopoles in the infinite volume 
limit. Then the finiteness of $b_c =na(\beta^n_c)$ suggests 
$\beta^n_c$ becomes infinite when the extendedness $n$ 
goes to infinity. $SU(2)$ lattice QCD is always (for all $\beta$) 
in the monopole condensed and then in the color confinement phase.\cite{thooft} This is one of what one wants to prove in the framework of lattice QCD.

Notice again that considering extended monopoles corresponds to performing 
a block spin transformation on the dual lattice. 
The above fact that the effective actions 
for all extended monopoles considered 
are the same for fixed $b$ means that the action may be  
the renormalized trajectory on which one can take the continuum limit.
See Fig.8.
Our results suggest the continuum monopole action takes the form
 predicted by Smit and Sijs.\cite{smit}

\section{Simulation of monopole action}
To test if the fixed monopole action is the renormalized trajectory, we have 
tried to simulate the system with the action.\cite{suzu95a}

We have obtained the following results:
\begin{enumerate}
\item
The action with quadratic interactions alone is not good, since it gives 
a longer monopole loop and a larger string tension than expected.
We need a term which gives 
a repulsive force 
between monopole currents.
\item
With such small repulsive terms, 
we get the nice fit of the loop length 
and the string tension on small lattices. 
Finite-size effects are seen to be small. 
\item
We have also made Monte-Carlo simulations in the $T\neq 0$ QCD.
Only the self-coupling term can reproduce the qualitative features of 
the deconfinement transition as seen in Fig.10. 
\end{enumerate}

\begin{figure}[tb]
\vspace{-1cm}
\epsfxsize=6cm
\begin{flushleft}
\leavevmode
\epsfbox{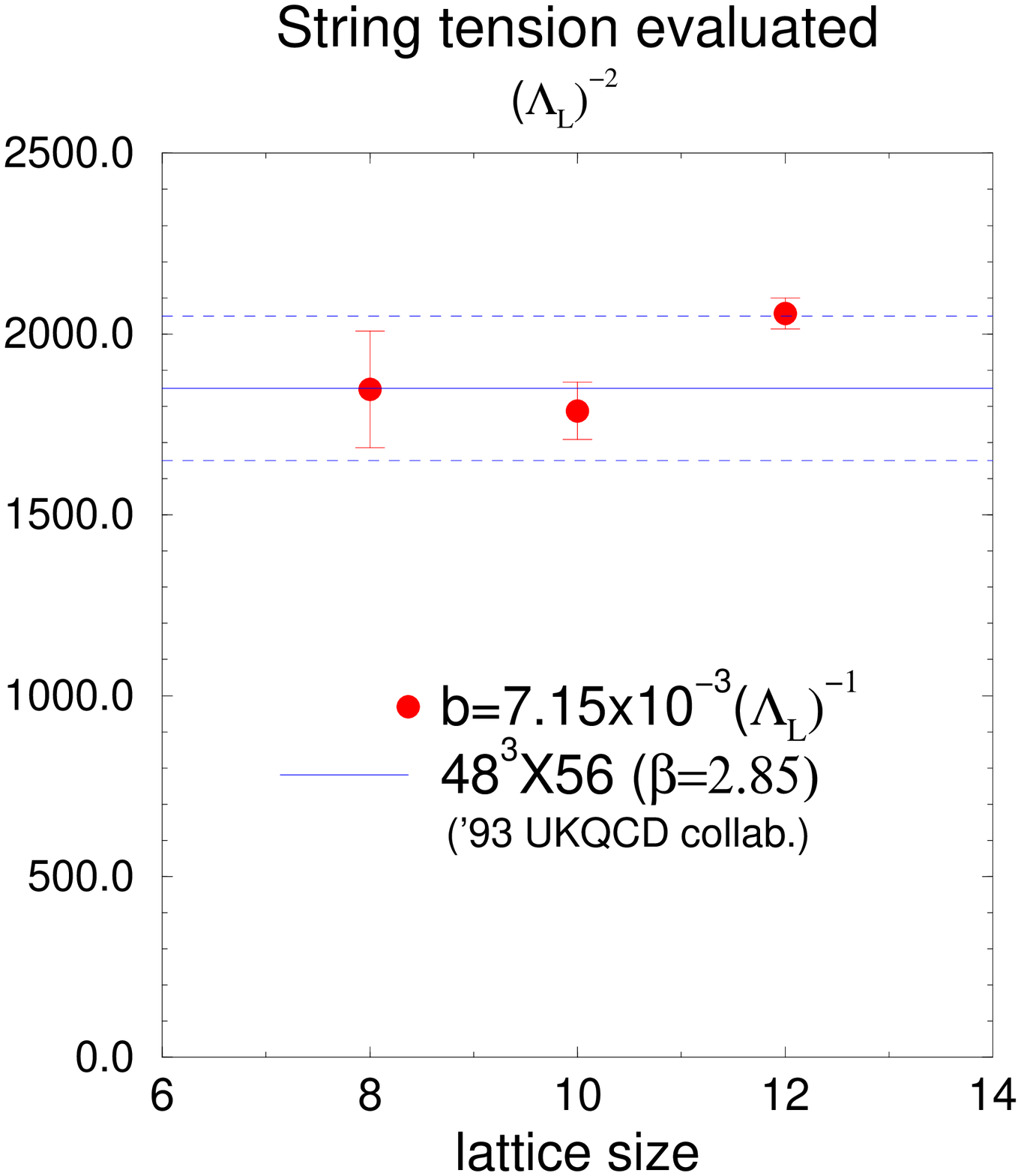}
\vspace{-1cm}
\end{flushleft}
\begin{flushleft}
Fig.9 \ The string tension versus\\
 the lattice size
\end{flushleft}
\label{f11}

\vspace{-9.5cm}

\epsfxsize=6cm
\begin{flushright}
\leavevmode
\epsfbox{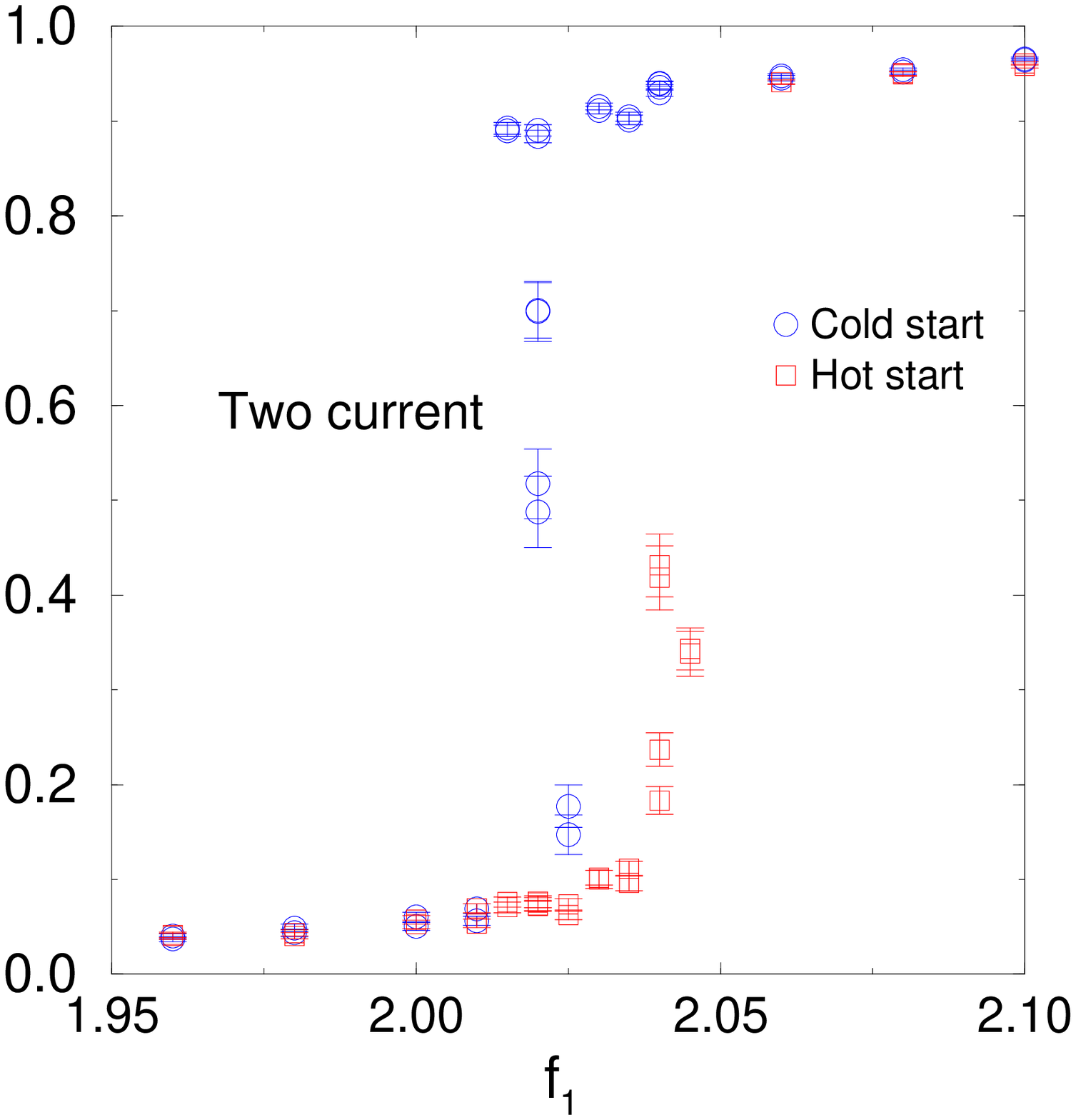}
\vspace{-1cm}
\end{flushright}
\begin{flushright}
\parbox{6cm}{
\begin{flushleft}
Fig.10 \ Polyakov loop from the self-coupling  monopole interactions  alone
in the two-current case
\end{flushleft}
}
\end{flushright}
\label{f13}
\end{figure}

\section{Discussions and outlook}

Some comments are in order.
\begin{enumerate}
\item
It is important to find an order parameter of confinement due to 
monopole condensation.\cite{suzu95a} 
The number of Dirac string $n_{i4}(s)$ may be a candidate.\cite{suzu95a}
$n_{i4}(s)$ is a magnetic $U(1)$ variant quantity.\cite{jevicki}
\item
Gauge independence should be proved if the monopole condensation 
is the real confinement mechanism.
The data of Polyakov loops from monopoles are encouraging.
That the correct string tension could be derived both in MA gauge 
and Polyakov gauge is also encouraging.
Gauge independent results will be obtained if we go to 
large $\beta$ on larger lattice. 
\item
How to test the correctness of this idea?
The theory 
predicts the existence of 
an axial vector glueball-like state  C($J^{pc}=1^{+-}$)
and 
a scalar glueball-like state $\chi$($J^{pc}=0^{++}$).
The masses seem to satisfy $m_c\sim m_{\chi}$.\cite{singh,matsu93,hay93}
The masses could not be too heavy. They have to exist under 2Gev.
To evaluate the correlation between the state and the light hadrons
in MC simulations of full QCD is very important to derive the
total width and the branching ratios.
\end{enumerate}

The author is thankful to Y.Matsubara, S.Kitahara, H.Shiba, S.Ejiri 
and \\ 
O.Miyamura for (partial) collaboration and fruitful discussions.
This work is financially supported by JSPS Grant-in Aid for 
Scientific  Research (B) (No.06452028).

\end{document}